# TrialChain: A Blockchain-Based Platform to Validate Data Integrity in Large, Biomedical Research Studies

Hao Dai[1†], H Patrick Young PhD[2,3†], Thomas JS Durant MD[2,4], Guannan Gong MS[5], Mingming Kang PhD[1], Harlan M Krumholz MD SM[2,3,5], Wade L Schulz MD PhD[2,4Δ], Lixin Jiang MD PhD[1Δ*]

[1]National Center for Cardiovascular Disease, Fuwai Hospital, Beijing, China; [2]Yale New Haven Hospital, Center for Outcomes Research and Evaluation, New Haven, CT; [3]Yale University School of Medicine, Department of Internal Medicine, Cardiology, New Haven, CT; [4]Yale University School of Medicine, Department of Laboratory Medicine, New Haven, CT; [5]Yale University School of Public Health, Department of Health Policy and Management, New Haven, CT

[†] These authors contributed equally to the manuscript.
[Δ] These are joint senior authors.

*To whom correspondence should be addressed: National Clinical Research Center of Cardiovascular Diseases, State Key Laboratory of Cardiovascular Disease, Fuwai Hospital, National Center for Cardiovascular Diseases, Chinese Academy of Medical Sciences and Peking Union Medical College, Beijing, 100037, China. jiangl@fwoxford.org

## Abstract

The governance of data used for biomedical research and clinical trials is an important requirement for generating accurate results. To improve the visibility of data quality and analysis, we developed TrialChain, a blockchain-based platform that can be used to validate data integrity from large, biomedical research studies. We implemented a private blockchain using the MultiChain platform and integrated it with a data science platform deployed within a large research center. An administrative web application was built with Python to manage the platform, which was built with a microservice architecture using Docker. The TrialChain platform was integrated during data acquisition into our existing data science platform. Using NiFi, data were hashed and logged within the local blockchain infrastructure. To provide public validation, the local blockchain state was periodically synchronized to the public Ethereum network. The use of a combined private/public blockchain platform allows for both public validation of results while maintaining additional security and lower cost for blockchain transactions. Original data and modifications due to downstream analysis can be logged within TrialChain and data assets or results can be rapidly validated when needed using API calls to the platform. The TrialChain platform provides a data governance solution to audit the acquisition and analysis of biomedical research data. The platform provides cryptographic assurance of data authenticity and can also be used to document data analysis.

## 1. Introduction

Data integrity and governance are key requirements for any data management platform. Within biomedical research and clinical trials, assurance of data validity and documentation of analytic steps are critical for translating results to high quality clinical care. Data manipulation, whether unintentional or due to scientific fraud [1-8], can be difficult to identify without a robust infrastructure for managing data assets. While improving the transparency of data acquisition and analysis requires a multifaceted approach [9-16], the use of emerging cryptographic technologies, such as blockchain, may reduce the risk of data manipulation and boost the confidence in conclusions made by the scientific community.

Software applications for the management and governance of biomedical research data are commercially available. However, these laboratory information management systems (LIMS) and clinical trial management systems (CTMS) are typically complex, internally managed systems that are not shared among collaborating investigators or sponsoring organizations. Furthermore, because these systems are privately managed, they are potentially a weak safeguard against misconduct. In addition, because of the complexity and expense of these systems, they are used in only a limited number of studies [17-20]. Novel approaches are therefore needed to improve the robustness and transparency of data governance systems as well as expand access to their capabilities.

Similar issues with stakeholder trust and data integrity have been identified in many industries [21-24]. Blockchain, a distributed ledger technology originally developed in 2008 for the Bitcoin platform [25], is a potential solution to the problem of trust in numerous use cases [26 27]. Several organizations, particularly in the financial industry, have started to explore whether blockchain technology can be successfully integrated into existing software to address the need for a more visible and immutable audit log [28]. More recently, the use of blockchain for applications outside of currency and financial services has also received significant attention. Within healthcare, blockchain has been proposed as a possible solution for managing patient and provider identity, permissions to healthcare data, and to manage participant consent [29-31].

Since 2014, the National Center for Cardiovascular Disease (NCCD), China in Beijing, has enrolled over 2 million participants as part of the China PEACE Million Persons Project (MPP) [32]. As a large, distributed study, our goal was to implement a platform that could ensure that data integrity and provenance could be verified and provide an audit trail of data modifications from the time of collection through analysis. The MPP data set includes a vast number of structured and unstructured data types, such as survey responses, medical records, echocardiogram reports, imaging studies, and genomic sequencing. Because of the number and diverse types of data elements, a strong data governance strategy, which includes both data management policies and technical safeguards, is especially important.

To create a robust data management environment, we implemented a blockchain-based platform to track data assets obtained for the study. The technical basis of blockchain is that each entry to the ledger is added to a growing list of blocks which are cryptographically signed with the current timestamp and a hash of the previous block. Clients, or miners, can then cryptographically validate each transaction to guarantee that data within the ledger has not been subsequently modified. While it is possible for actors to intentionally fabricate data at the site of capture before sharing or storing it, this approach helps guarantee that data cannot be maliciously modified after the fact or in bulk to fit a desired outcome, and also provides a shared ledger that can be viewed and validated by multiple parties. Here, we present the technical implementation of the NCCD Data Science Platform (NDSP), which includes a blockchain-based application used to maintain a cryptographically-secured ledger of data assets.

## 2. Methods

### 2.1 Data Platform and Data Acquisition

The TrialChain application was integrated into the NDSP, a data science platform developed to provide data management for the Million Persons Project. The NDSP core computing platform consists of a Hadoop cluster for distributed data management and a high-performance computing (HPC) cluster for primary and secondary next-generation sequencing (NGS) analysis. Hadoop was deployed to a mixed operating system (OS) environment with nodes running either CentOS7 or Ubuntu 14.04 (GPU-enabled nodes). Core Hadoop applications including the Hadoop Distributed File System (HDFS, v 2.7.3), Yet Another Resource Negotiator (YARN, v 2.7.3), Spark (v 1.6.2), Hive (v 1.2.1), and Sqoop (v 1.4.6) were deployed following best practices for high-availability and secured with Kerberos (Figure 1). The Apache NiFi software (v 0.4.0) was also deployed to an edge node for data ingestion and stream processing. Docker was deployed to all edge and worker nodes to support a microservice architecture for ancillary and analytic applications. The HPC was integrated with Hadoop client services and the SLURM resource manager deployed on CentOS7.

### 2.2 TrialChain Infrastructure and Libraries

The TrialChain platform was developed with a microservice architecture using Docker (v 18.03.1-ce) and deployed with Docker Compose (v 1.21.0), based on the Ubuntu 16.04 base Docker image (http://github.com/ComputationalHealth/TrialChain). The open source MultiChain platform (v 1.0.2) was used for the private blockchain infrastructure. The TrialChain software was developed in Python (v 3.6.2) with the Savoir (v 1.0.6), Web3.py (v 3.16.5), Ethereum (v 2.3.1), and gevent (v 1.2.2) libraries. The Falcon (v 1.4.1) and Flask (v 0.12.2) frameworks were used for the application programming interface (API) and administrative web interface, respectively. The MultiChain Explorer (v 0.8) package was used to provide a user interface for introspecting the private blockchain. A Docker container running Geth (v 1.8.0), the Go implementation of the Ethereum protocol, was also deployed to allow for a local cache and link to the public Ethereum network.

## 3. Results

### 3.1 Private Blockchain Architecture

MultiChain, a private blockchain, was chosen to log data collected within the platform, as it offers several advantages over public blockchains, as described in detail in the discussion. One such advantage is the ability to approve nodes that join the network and apply specific permissions to these nodes. As such, the private blockchain network is initiated by an initial administrative "master" node that creates the new blockchain, defines the chain's parameters, and grants permissions to additional "client" nodes. MultiChain can expose a remote procedure call (RPC) endpoint that can be used to interact with the network and the master node was configured to allow RPC connections from other TrialChain components within a shared Docker network (Figure 2A).

MultiChain client nodes were deployed to create a distributed, multi-host architecture. A client node was created for each TrialChain component, including the TrialChain web service and web portal, and were configured to accept RPC connections from the associated TrialChain component on a local Docker network. These nodes were then given connect, send, and receive permissions from the initial master/administrative node and allowed to join the network. This allows for each application to run independently and replicates the blockchain data to each host to protect from data loss in the event of node failure. All transactions, including adding or removing nodes or changing node permissions, are logged within the blockchain.

### 3.2 Private-Public Blockchain Integration

While the private blockchain cryptographically links each transaction within the MultiChain environment, private blockchains have potential limitations related to data immutability. To provide additional cryptographic assurance and public validation of the private blockchain data, we integrated a public blockchain by periodically submitting the state of the private chain (latest blockhash) as a transaction on the public Ethereum network (Figure 2B). To achieve this private-public integration, we deployed a local Ethereum node to synchronize and allow for programmatic interaction with the public blockchain. To provide an interface to manage the private blockchain and the public blockchain integration, we implemented an administrative web portal.

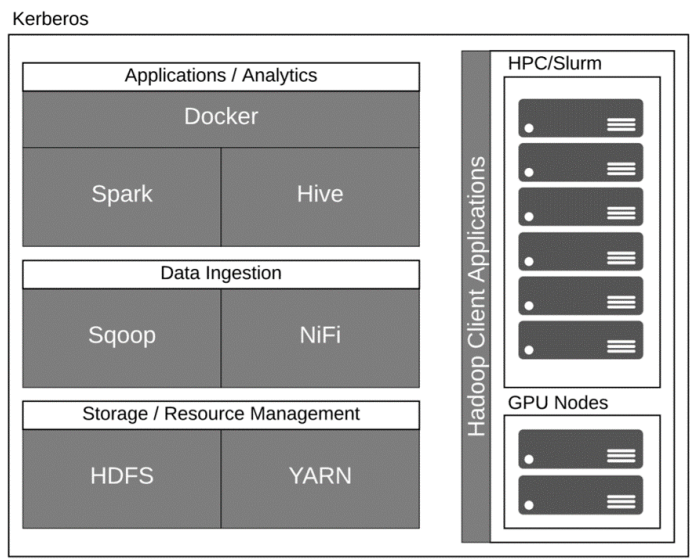

**Figure 1:** The NCCD Data Science Platform includes several core components from Hadoop for data storage, ingestion, and analysis. Other core infrastructure components, such as Apache Zookeeper and Ambari, are also used to support the platform. A high-performance cluster and GPU-enabled nodes for data analysis were integrated into the platform through installation of the Hadoop client applications.

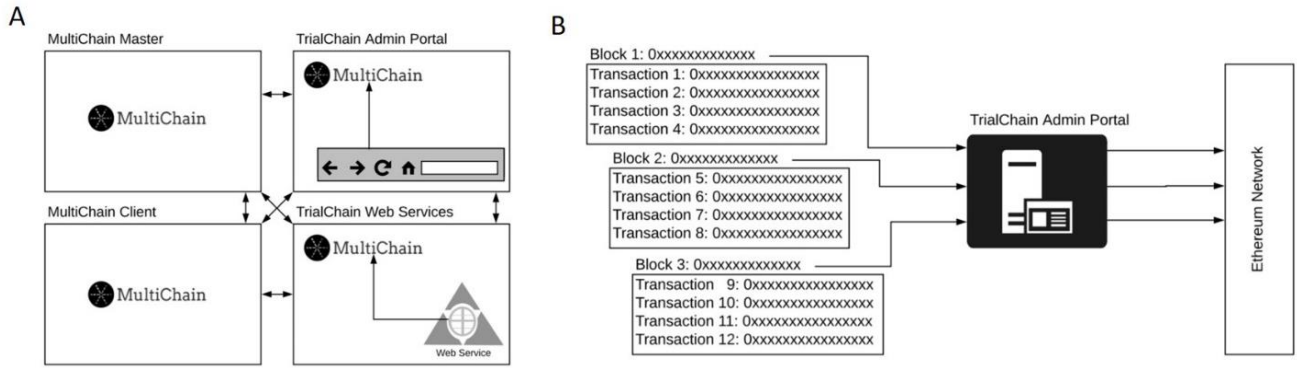

**Figure 2:** TrialChain workflow and integration with the public Ethereum network. A) A master node is initially deployed to create the blockchain and assign initial permissions. Subsequently, additional client nodes can be added to the network either locally or, with proper permissions and network access, at remote locations. TrialChain services are deployed within individual Docker containers along with MutliChain client nodes to provide local access and data replication. B) The TrialChain administrative portal periodically synchronizes the most recent blockhash, which represents the current chain state, with the public Ethereum network.

The backend of the web portal was developed to handle constructing and submitting raw transactions to the Ethererum network. An Ethereum wallet was created specifically for the application, and its private key was exposed to the web portal backend through the environment variables of the Docker container. Although no Ether is moved in the validation transactions, the transaction still carries a small cost to incentivize miners on the public network to include it in the proof-of-work, so a small but non-zero quantity of Ether must be maintained in the administrative account's wallet.

The web portal defaults to using the local Ethereum node to submit transactions to the network, but if the local chaindata is not in-sync, transactions may also be directed through public third-party APIs such as Infura or BlockCypher. Transactions may be scheduled, or an admin user of the web portal can manually initiate a validation transaction on the Ethereum network at any time and can specify a different external API to use through a webform. A dashboard displays the status of the local chaindata as well as the dollar-equivalent balance of ether remaining in the wallet and the current estimated cost of a validation transaction on the Ethereum network.

When the private-public validation task is initiated, the web app backend establishes a connection with the private blockchain and retrieves the latest confirmed blockhash. A connection is then established with the public Ethereum network through either the provided external URL or the local Geth node's exposed JSON-RPC API port. A transaction containing the local blockhash and any additional metadata is then constructed and signed with the local private key (Figure 3A). The transaction costs are estimated before submitting the transaction and if the account balance is too low, a connection cannot be established, or the transaction request does not return a transaction hash response, the user is notified, and the transaction is canceled. If the transaction submission is successful, the web app records the current time, the submitted blockhash, the transaction hash, and the account address of the wallet in a local postgres database supporting the web application.

*3.3 TrialChain and NDSP Integration*

To capture data assets entering the NDSP, a REST-based API based on the Falcon library was created to provide a secure method that could be used to submit data to TrialChain from multiple applications or data sources. Data assets can be submitted to this API with a valid POST request originating from any authorized IP address. Since a majority of data ingestion for this platform was performed with the Apache NiFi software, we created a custom NiFi workflow to hash and submit the data assets to the TrialChain platform. New data assets acquired from a number of data sources, including a HPC used for genetic data analysis, are retrieved with NiFi and both an md5 and SHA256 hash are generated. The md5 hash is used as the index ID for the MultiChain Asset class when it is issued, as this index is limited to 32 characters. The SHA256 hash is recorded as an attribute of the data asset to provide further assurance of the data asset, since the md5 hashing algorithm is not considered cryptographically secure. NiFi is used to add additional metadata to a JSON message which is then sent as a POST request to the TrialChain API (Figure 3B). The API then submits the necessary data asset information to a local MultiChain node and the data is broadcasted to the entire network.

*3.4 Retrospective Data Validation*

To determine whether a data asset has been verified in the blockchain, along with the initial time of validation, an interface to query and review the status of assets on the blockchain was added to the administrative web portal. This allows users to verify that data existed in a specific state at the specified time, to demonstrate that data have not been changed since acquisition. When a user submits an md5 hash entered in a webform, the application queries the local MultiChain node to identify the asset ID that matches the md5 hash. If none is found, the user is notified; otherwise, the web application determines at what time the local blockchain was subsequently synchronized with the public Ethereum network and confirms the status of that transaction. The returned metadata includes the transaction receipt, the number of confirmations and a hyperlink to the transaction report on the etherscan.io website, a publicly available portal for browsing Ethereum transactions. An embedded object on the dashboard of the web application also exposes the MultiChain explorer so that local MultiChain transactions can also be viewed. The same functionality was also implemented as a local Python application to allow data assets to be hashed locally and verified through an API call to the web portal. After generating an md5 and SHA256 hash for the provided file, the same information provided in the web portal is returned as a JSON message to the client and displayed to the user through the command line interface (Figure 3C).

A
```
{
    "data": "0x303...038",
    "gasprice": 4000000000,
    "hash": "0x69f...3bc",
    "nonce": 19,
    "r": 349...847,
    "s": 501...336,
    "sender": "0x92b...ed1",
    "startgas": 55112,
    "to": "0x92b...ed1",
    "v": 28,
    "value": 0
}
```

B
```
{
    "hash.md5": "b5a5dfa95146b28cd79e791fcd4eaace",
    "hash.sha256": "b3e...5be",
    "processed.ts": 1519316242073,
    "source.uri": "network.trialchain/nifi/getfile/mnt/data"
}
```

C
```
{
    "asset": "abd992b7a76a6ae845752ac8d1f72812",
    "confirmations": "17909",
    "ethStatus": "Confirmed",
    "ethTxId": "0x6bc...dcf",
    "issueTxId": "6ca...f75",
    "issued": "Thu, 05 Apr 2018 20:09:38 GMT",
    "multiChainHash": "00f...a0e",
    "sha256": "d3b...177",
    "source": "network.trialchain/nifi/getfile/mnt/data",
    "validated": "Fri, 06 Apr 2018 20:38:27 GMT"
}
```

**Figure 3:** TrialChain data models for web service integration. A) The TrialChain administrative portal creates a transaction containing the required Ethereum parameters along with the 'data' field, which contains the most recent local blockhash. The transaction is signed with the local private key and submitted to the network. B) Data assets are hashed within NiFi and a JSON message containing the file hashes, a timestamp, unique source identifier, and any additional metadata is submitted to the TrialChain web service to create a local transaction. C) After submitting the hash of a file that is already present in the blockchain, TrialChain returns a JSON response with related transaction identifiers, synchronization status with the Ethereum network, and related timestamps.

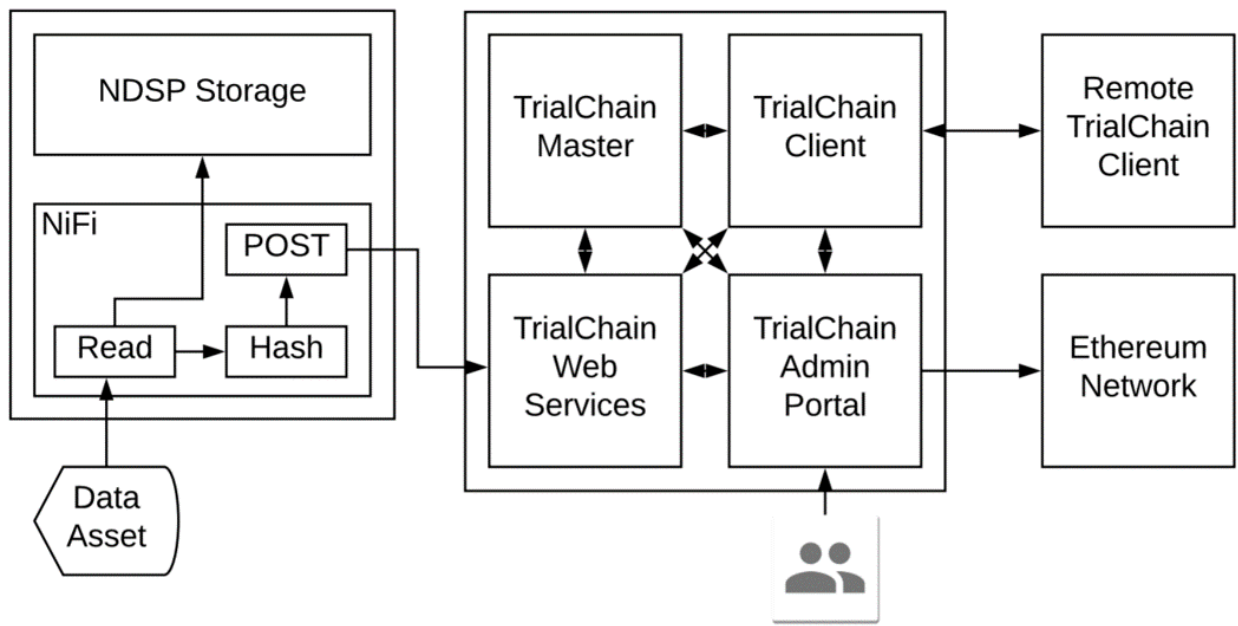

**Figure 4:** TrialChain architecture and workflow. After acquiring data assets, the file hashes are submitted to the TrialChain platform and files are stored within the NDSP. Assets are created within the local blockchain and periodically synchronized to the public Ethereum network. Users can verify the status of the network and file authenticity through the administrative portal and remote clients can be configured to allow for further data distribution.

## 4. Discussion

Through the use of a distributed, cryptographic ledger, the TrialChain platform (Figure 4) is able to provide a publicly-verifiable audit trail of data assets. With the continued growth and importance of digital data for clinical trials and other biomedical research, novel approaches such as demonstrated here will allow for greater confidence in the authenticity of the data used to generate results. In addition, the integration of private and public blockchain networks offers an approach that can be used to decrease the cost of running blockchain-based solutions, while still offering a high-level of transparency and authenticity.

Like Bitcoin and many other major blockchain implementations, MultiChain consists of a network of nodes that validate shared transaction histories through a proof-of-work algorithm [33]. In contrast to public blockchains, only client nodes that are expressly known to the master node are permitted to participate in mining or to access the chain data. There are several advantages to using a private blockchain platform such as MultiChain. The software required to run MultiChain, including the private blockchain itself, can be locked to a specific version of the MultiChain protocol and therefore does not necessarily require upgrades to its supporting components as would be required to keep pace with development efforts in public blockchains. Also, transaction costs can be set when configuring the blockchain, and can therefore be free of any monetary or cryptocurrency charge besides the cost of maintaining the computing resources and, because transactions on the network are exclusively for data governance of PEACE and MPP data assets, the resources required for rapid transaction uptake and block validation is achievable with standard commercial desktops and small virtual machines. Furthermore, as the blockchain and any metadata associated with the data assets are never disseminated outside of the private network, the use of a private chain provides an environment to log more sensitive data than would be appropriate to be shared on a public network.

While private blockchains have many advantages, because of the small size of the network as well as the lack of diversity of incentives among participating nodes, private blockchains are vulnerable to mutability through collusion among the nodes to either roll back the chain to an earlier state or exclude certain transactions from inclusion in a block. Although technically difficult, it would be both theoretically possible and practically feasible to alter the data verified by this chain after it was initially submitted [33 34]. To guarantee that the private chain is trustworthy, the latest blockhash in the private chain is periodically (once per day) shared with the Ethereum network as a transaction on the public ledger. As it is practically infeasible to corrupt the public network, it is possible for all stakeholders to verify that every data asset in the private chain has been unchanged since the nearest synchronization date since its upload.

For the NDSP, both master and client nodes are running within a single institution, however the platform architecture can just as easily support nodes installed on external networks, such as in systems held by third-party stakeholders. Including additional stakeholders with differing incentives as nodes in the network therefore provides a concrete mechanism to build trust among collaborators while simultaneously making misconduct by individual stakeholders more difficult. The distribution of nodes also allows stakeholders to maintain a local copy of the transaction log for future review or analysis.

Since the TrialChain API can accept a web service request from multiple sources, hashes can be submitted not only at the time of data ingestion, but also for subsequent analyses or modification. For example, hashes can be submitted and assets created for genomic data pre- and post-alignment, with the latter referencing the first asset within the private blockchain. This approach allows not only for assurance of data authenticity, but also an audit trail of analyses and modifications. The dynamic metadata model also allows for assets to contain an evolving set of information throughout the course of a study.

Given the high value of biomedical data and need for verifiability of data assets in clinical trials, blockchain technology provides a reasonable solution that can be rapidly integrated into existing infrastructure to provide a publicly-verifiable audit log. As the number and size of data assets continues to increase, the risk of data manipulation, either intentional or accidental, can be mitigated with technology safeguards, such as blockchain. In particular, the longitudinal aspect of blockchain provides a robust and automated layer of protection against retrospective or batch data manipulation intended to validate a given hypothesis. While not a solution to all potential causes of data manipulation, such as fabrication of data at the time and place of capture, implementing these technological safeguards can provide rapid assurance to study sponsors and the public that data are being used appropriately to drive discovery.

**Funding and Acknowledgements**

This project was partially supported by the CAMS Innovation Fund for Medical Science (2017-I2M-2-002, 2016-I2M-1-006, 2016-I2M-2-004); the Ministry of Finance of China and National Health and Family Planning Commission of China; the China-WHO Biennial Collaborative Projects 2016-2017 (2016/664424-0); the National Key Technology R&D Program (2015BAI12B01, 2015BAI12B02); Research Special Fund for Public Welfare Industry of Health (201502009); the 111 Project from the Ministry of Education of China (B16005); and the PUMC Youth Fund and the Fundamental Research Funds for the Central Universities (2017330003).

**Competing Interests**

*Dr. Krumholz* is a recipient of research agreements from Medtronic and from Johnson & Johnson (Janssen), through Yale, to develop methods of clinical trial data sharing; is the recipient of a grant from the Food and Drug Administration and Medtronic, through Yale, to develop methods for post-market surveillance of medical devices; works under contract with the Centers for Medicare & Medicaid Services to develop and maintain performance measures; chairs a cardiac scientific advisory board for UnitedHealth; is a participant/participant representative of the IBM Watson Health Life Sciences Board; is a member of the Advisory Board for Element Science and the Physician Advisory Board for Aetna; and is the founder of Hugo, a personal health information platform. *Dr. Schulz* is a consultant for Hugo, a personal health information platform.